\DeclareUnicodeCharacter{2212}{-}
\documentclass[conference]{IEEEtran}
\IEEEoverridecommandlockouts
\usepackage{hyperref}
\usepackage[capposition=top]{floatrow}
\usepackage[pdftex]{graphicx}
\usepackage[utf8]{inputenc}
\usepackage{amsmath}
\usepackage{algorithmic}
\usepackage{array}

\usepackage{stfloats}
\usepackage{url}
\usepackage[numbers]{natbib}

\makeatletter
\newcommand{\linebreakand}{%
  \end{@IEEEauthorhalign}
  \hfill\mbox{}\par
  \mbox{}\hfill\begin{@IEEEauthorhalign}
}
\makeatother

\begin{document}

\title{A Data Science Pipeline for Algorithmic Trading: A Comparative Study of Applications for Finance and Cryptoeconomics
\thanks{Tianyu Wu, Saad Lahrichi, Jiayi Li, and Carlos-Gustavo Salas-Flores were supported by the Summer Research Scholar Program at Duke Kunshan University as research affiliates in Prof. Luyao Zhang's project entitled "How Fintech Empowers Asset Valuation: Theory and Applications" }
}

\author{\IEEEauthorblockN{1\textsuperscript{st} Luyao Zhang}
\IEEEauthorblockA{\textit{Data Science Research Center and Social Science Division } \\
\textit{Duke Kunshan University}\\
Suzhou, China \\
lz183@duke.edu}
\linebreakand

\and
\IEEEauthorblockN{1\textsuperscript{st} Tianyu Wu}
\IEEEauthorblockA{
\textit{Duke Kunshan University}\\
Suzhou, China}
\and
\IEEEauthorblockN{1\textsuperscript{st} Saad Lahrichi}
\IEEEauthorblockA{
\textit{Duke Kunshan University}\\
Suzhou, China}
\and
\IEEEauthorblockN{2\textsuperscript{nd}Carlos-Gustavo Salas-Flores}
\IEEEauthorblockA{
\textit{Duke Kunshan University}\\
Suzhou, China}
\and
\IEEEauthorblockN{2\textsuperscript{nd} Jiayi Li}
\IEEEauthorblockA{
\textit{Duke Kunshan University}\\
Suzhou, China}
}

\maketitle

\begin{abstract}
Recent advances in Artificial Intelligence (AI) have made algorithmic trading play a central role in finance. However, current research and applications are disconnected information islands. We propose a generally applicable pipeline for designing, programming, and evaluating the algorithmic trading of stock and crypto assets. Moreover, we demonstrate how our data science pipeline works with respect to four conventional algorithms: the moving average crossover, volume-weighted average price, sentiment analysis, and statistical arbitrage algorithms. Our study offers a systematic way to program, evaluate, and compare different trading strategies. Furthermore, we implement our algorithms through object-oriented programming in Python3, which serves as open-source software for future academic research and applications.
\end{abstract}


\section{Introduction}
Recent advances in AI have made algorithmic trading (AT) play a central role in finance:~\citet{burgess_2019_an} estimates that AT accounted for 80\% of the entire equity turnover in the U.S. by transaction volume in 2018, up from 50\% in 2011. AT empowers traders with cutting-edge technological innovation so that they can execute trades with predefined strategies, capturing profitable opportunities faster and with a low attention cost~\cite{virgilio_2019_highfrequency}. Moreover, AT contributes to market efficiency and liquidity~\cite{gomber_2018_algorithmic}.
However, current research and applications are disconnected information islands~\cite{reznik_highfrequency}. AT in financial industries is generally a black box for which scientific evaluation is almost impossible~\cite{azzutti_2021_machine}. Academic research has more clearly elaborated on algorithms. However, current research diverges in many ways. The lack of process consistency makes ceteris paribus comparison difficult. Moreover, there is little open-source software for designing AT strategies, which leads to unnecessary obstacles to collaborative learning, research, and innovation. 
\par

\begin{figure}[!htpb]
\centering
\includegraphics[width=3.6in]{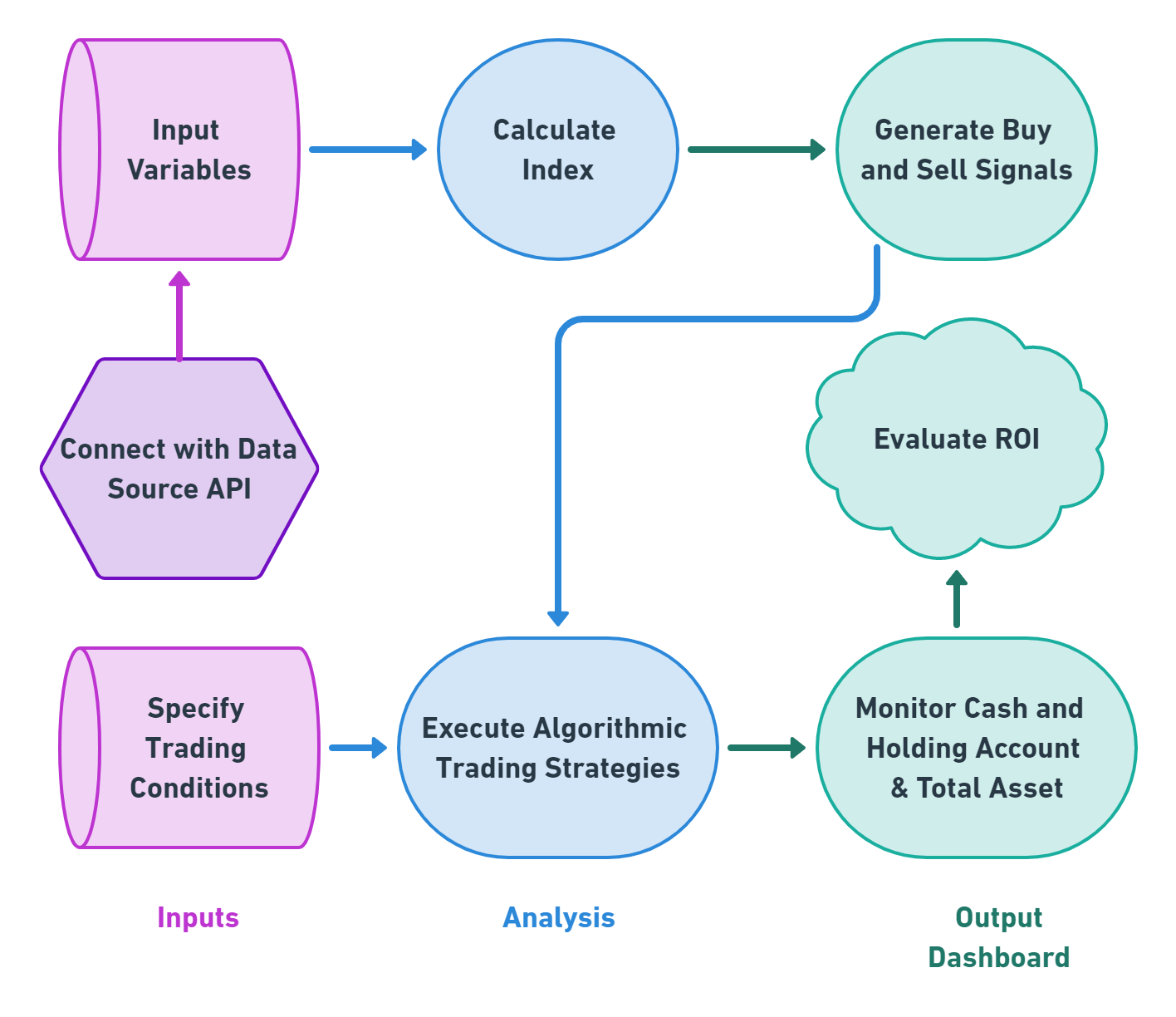}

\caption{\textbf{The data science pipeline for algorithmic trading}: The general workflow includes inputs (pink), analysis (blue), and the output dashboard (green). At the input stage, we first connect with the data source API to input variables for calculating indices that are necessary to calculate buy-and-sell signals and then specify trading conditions. At the analysis stage, we calculate indices that are necessary to calculate buy-and-sell signals and execute algorithmic trading strategies. At the output stage, we visualize three dashboards: (1) the time series of buy-and-sell signals, (2) the cash and holding accounts and total assets, and (3) the return on investment (ROI). }

\label{fig:1}
\end{figure}

We propose a generally applicable pipeline for designing, programming, and evaluating the algorithmic trading of stock and crypto tokens. Figure~\ref{fig:1} represents the general workflow: inputs, analysis, and the output dashboard. We first connect with the data source API to input variables for calculating indices that are necessary to calculate buy-and-sell signals; then, we further specify trading conditions to execute algorithmic trading strategies based on the signals. Finally, we generate visualizations to monitor cash and holding accounts and evaluate the return on investment (ROI). We demonstrate that our data science pipeline is generally applicable to conventional algorithms, including the moving average crossover, volume-weighted average price, sentiment analysis, and statistical arbitrage algorithms. 

Our study offers a systematic way to program and compare different trading strategies. Furthermore, we implement our algorithms by object-oriented programming in Python3, which serves as open-source software for future academic research and applications. We introduce the methodology applied to two algorithms in Section II, present the data and results in Section III, and discuss the results in Section IV. 

\section{Methodology}
\subsection{Moving Average Crossover}

The simple moving average (SMA), first proposed by Joseph E. Granville~\citet{josephensigngranville_1963_new}, is an indicator of the trend of stock prices~\cite{raudys_2018_optimising,bhattacharjee_2019_stock}.  As in Equation~\ref{eq:1}, the SMA with window $n$ is the average of the stock price in the past $n$ days.

\begin{equation}\label{eq:1}
    \text{SMA}_{t}^{n} = \frac{1}{n} \sum_{i=t-n+1}^{t}p_i \\
\end{equation}

A moving average crossover occurs when a short-window moving average (SWMA) and a long-window moving average (LWMA) intersect. On the one hand, the SWMA is an indicator of recent market prices. On the other hand, the LWMA represents the long-term or equilibrium price. A buy signal appears when the SWMA (such as the 50-day moving average) crosses above the LWMA (such as the 200-day moving average), predicting a bull market, and this is where the golden crossover occurs. The embedded idea is that as long-term indicators carry more weight, the golden crossover indicates a bull market on the horizon and is reinforced by high trading volumes. Similarly, a sell signal is released when the SWMA crosses below the LWMA, predicting a bear market, and this is where the death crossover happens. Some researchers have already tested this trading rule on some commonly used indices in the US stock market, finding that this technical analysis can generate a higher return on investment than the buy-and-hold strategy; however, it still cannot reflect a downturn in a timely manner due to its lag property~\cite{brock_1992_simple}.

\begin{figure}[!htpb]
\centering
\includegraphics[width=3.6in]{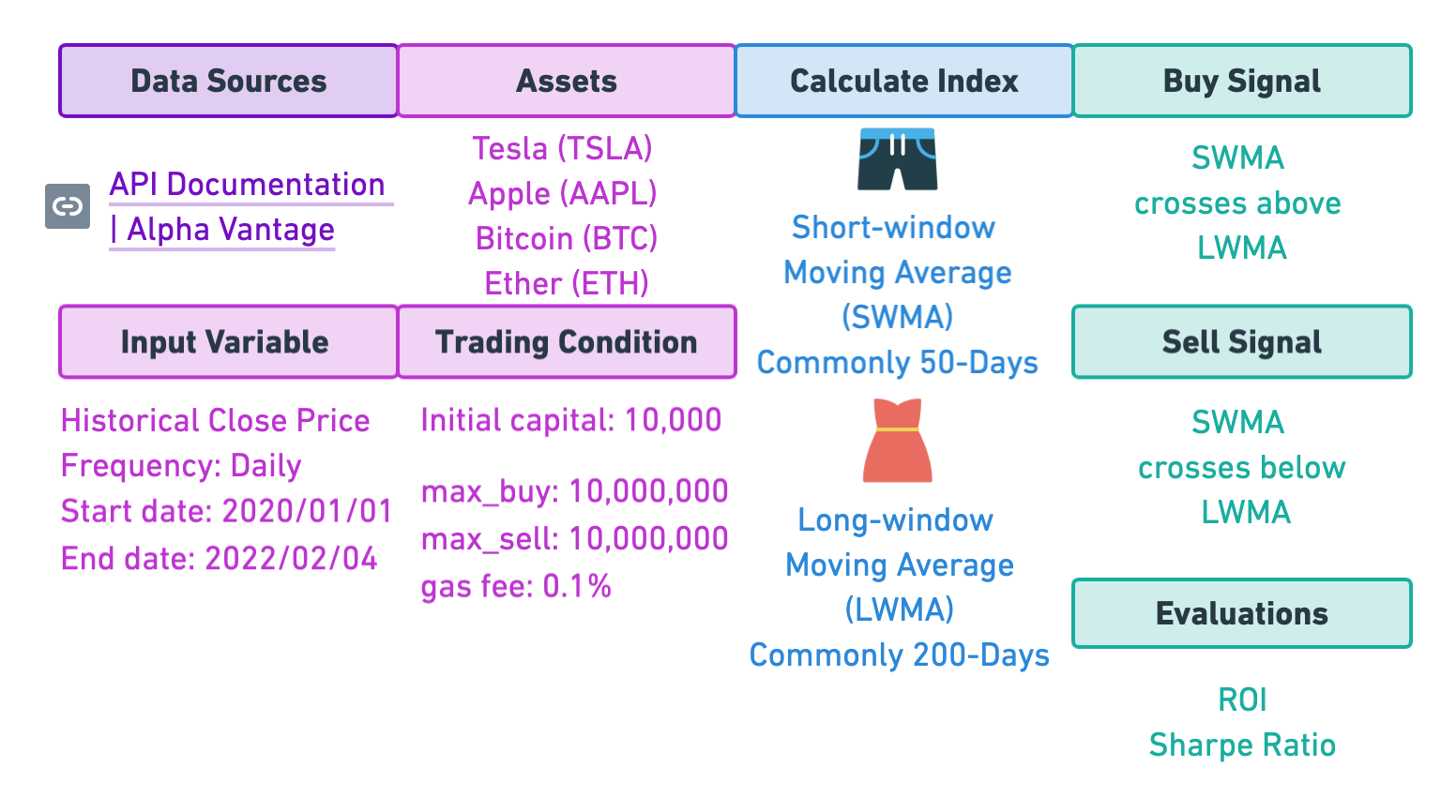}
\caption{\textbf{Moving average crossover}: The algorithm starts from the left with inputs (pink), from which we obtain our data from the Alpha Vantage API (purple). The raw input variables are historical daily closing prices. Then, we perform our moving average analysis (blue) to determine the buy-or-sell signals (green). Lastly, we evaluate the performance by using the ROI and Sharpe ratio as metrics.}

\label{fig:2_1}
\end{figure}

Figure~\ref{fig:2_1} represents the general pipeline for the moving average crossover strategy. We first input historical closing prices through the Alpha Vantage API, from which we can obtain all daily closing price data from the US stock market from the time to market, but we select 40\% of the historical data here because we would like to make a comparison based on the return on investment (RoI) and Sharpe ratio more intuitively in the Discussion section. Then, we calculate the SWMA and LWMA for generating buy-and-sell signals. Next, we simulate the strategies with an initial capital of 10,000 USD and set a maximum capacity to enable buying and selling at each crossover. Here, for simplicity, we ignore the transaction fee. Finally, we evaluate the performance of this strategy by comparing the ROI and Sharpe ratio to a buy-and-hold strategy.

\subsection{Volume-Weighted Average Price}

The SMA provides the recent price trend of a security. However, it does not take into account the level of volume traded at each price level. On the other hand, the volume-weighted average price (VWAP) gives the average price for intraday trading weighted by the transaction volume at each price. Equation (3) represents the formal definition of the VWAP, where P\_t is the price at time t and Q\_t is the corresponding volume traded at that price.

\begin{equation}\label{eq:2}
    VWAP = \frac{\sum{P_t \cdot Q_t}}{\sum{Q_t}} \\
\end{equation}

\citet{cesari_2012_effective} suggest that traders have incorporated institutional investors’ tendency into target the VWAP into their intraday trading strategies~\cite{cesari_2012_effective}. A bullish/bearish market is indicated when prices cross above/below the VWAP and traders shall buy/sell accordingly ~\cite{berkowitz_1988_the,admati_1988_a,andersen_1996_return}. Menkhoff (2010) indicates that the performance of VWAP strategies varies significantly and that the VWAP does not necessarily outperform the buy-and-hold strategy~\cite{menkhoff_2010_the}.

\begin{figure}[ht]
\centering
\includegraphics[width=3.6in]{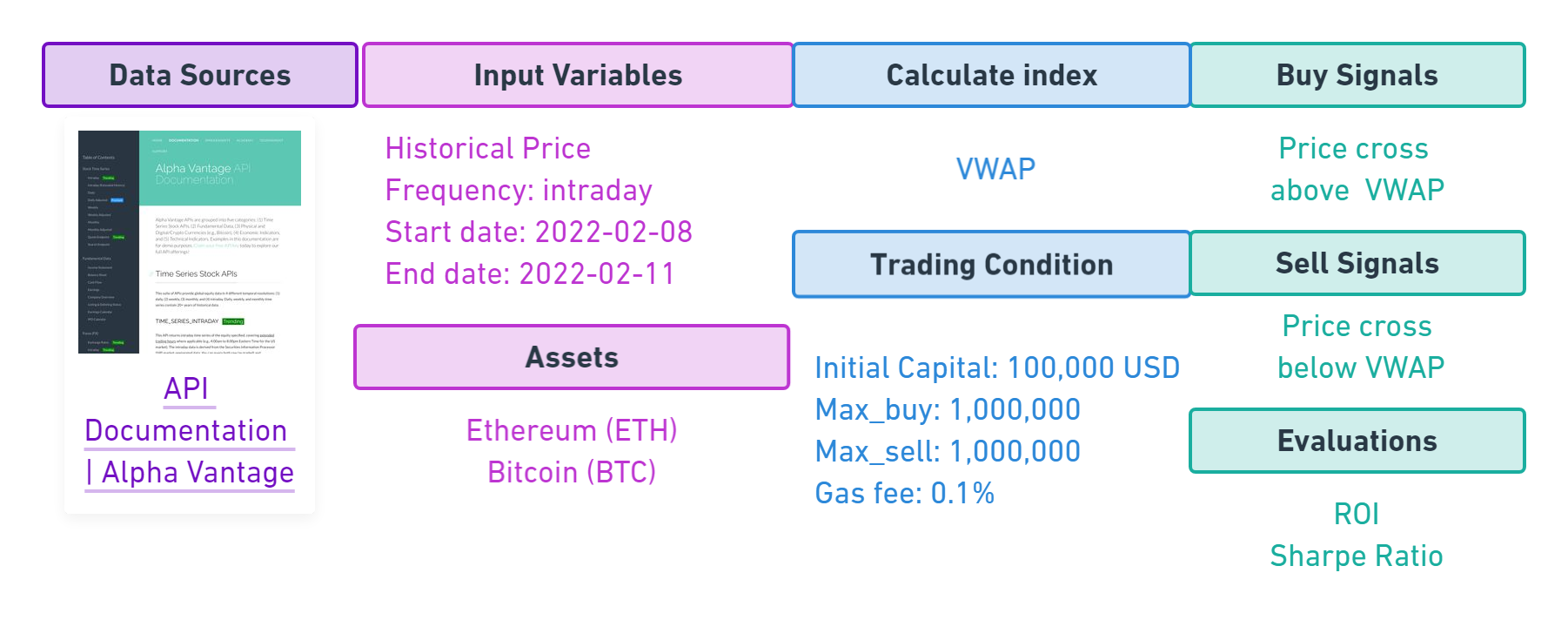}
\caption{\textbf{Summary of the VWAP trading algorithm}: The data sources used are shown in purple. We relied on the Alpha Vantage API to collect the necessary data. The input variables used and the assets we trade are in pink. The trading conditions and the calculated index (in our case, the VWAP) are shown in blue. Green represents the conditions for buy/sell signals as well as the metrics used for evaluating the algorithm.}
\label{fig:2_2}
\end{figure}

Figure~\ref{fig:2_2} represents the pipeline for a VWAP case study. We keep most parts consistent with the pipeline for the SMA. Unfortunately, the VWAP index is directly available from the Alpha Vantage API~\footnote{Alpha Vantage is a website that offers financial market data through its developer-friendly APIs. Alpha Vantage provides both traditional assets data and forex and cryptocurrency data. Upon installation, we used its time series stocks API to retrieve intraday time series of the stocks we explored. We also used the tech indicators API, from which we retrieved daily VWAP data for the same stocks. Without that API, we would have had to calculate the VWAP by applying its formula to the data. 
We have used Alpha Vantage because its API is among one of the few that offers free intraday data. Unlike the open-high-low-closing data that many websites offer, intraday data are rarer and usually are not free of charge. Quandl, for example, has high-quality intraday data, but they are not free. 
A key limitation of the use of Alpha Vantage’s free API is that it only allows us to query only the last 15 days.  It is worth noting that the documentation says that the TIME\_SERIES\_INTRADAY method can retrieve 1-2 months of intraday data, while the TIME\_SERIES\_INTRADAY\_EXTENDED returns the trailing 2 years. As another important limitation, the maximum number of requests is 5 API requests per minute and 500 API requests per day. Alpha Vantage offers a  premium API key, which allows for a higher call limit as well as more historical data.} only for traditional stocks. For cryptocurrencies, we perform the calculation ourselves. 

\section{Data and Results}
\subsection{Data}
\subsubsection{Moving Average Crossover Data}
Table \ref{table:1} shows all the data used in the moving average crossover algorithm:

\begin{table}[!htbp]
\centering
\resizebox{\textwidth}{!}{%
\begin{tabular}{|c|c|c|c|}
\hline
\textbf{Variable} & \textbf{Frequency} & \textbf{Unit} & \textbf{Description} \\ \hline
\textbf{Date} & daily & YYYY-MM-DD & \begin{tabular}[c]{@{}c@{}}Date and time for \\ which the data were recorded\end{tabular} \\ \hline
\textbf{Close} & daily & USD & \begin{tabular}[c]{@{}c@{}}Price at which the stock ended \\ trading in a given time period\end{tabular} \\ \hline
\textbf{Short MA} & daily & USD & \begin{tabular}[c]{@{}c@{}}Average price of a security within a \\ certain period, typically 50 days. \end{tabular} \\ \hline
\textbf{Long MA} & daily & USD & \begin{tabular}[c]{@{}c@{}}Average price of a security within a certain \\ period, typically 200 days. \end{tabular} \\ \hline
\textbf{Signal} & - & - & \begin{tabular}[c]{@{}c@{}} Buy-and-sell signal (e.g., TSLA, AAPL \\ for stock, BTC, ETH for crypto)\end{tabular} \\ \hline
\end{tabular}%
}
\caption{Moving Average Crossover: Data \\
Data Source: Alpha Vantage API. This table shows the raw data that we retrieved from the API. The price of stocks and crypto assets corresponds to the closing price of the given asset on a given date.}
\label{table:1}
\end{table}

\subsubsection{Volume-Weighted Average Price}
Using the Alpha Vantage API, we collected intraday stock data derived from securities information processor (SIP) market-aggregated data as well as intraday cryptocurrency data. The interval chosen was 5 min. Table \ref{table:vwap} summarizes the data. It shows the variables we use, the frequency at which these variables are collected, the unit of each variable, and a short description.

\begin{table}[!htbp]
\centering
\resizebox{\textwidth}{!}{%
\begin{tabular}{|c|c|c|c|}
\hline
\textbf{Variable} & \textbf{Frequency} & \textbf{Unit} & \textbf{Description} \\ \hline
\textbf{Date} & 5 min & \begin{tabular}[c]{@{}c@{}}YYYY-MM-DD \\ HH:MM:SS\end{tabular} & \begin{tabular}[c]{@{}c@{}}Date and time for \\ which the data were recorded\end{tabular} \\ \hline
\textbf{Close} & 5 min & USD & \begin{tabular}[c]{@{}c@{}}Price at which the stock ended \\ trading in a given time period\end{tabular} \\ \hline
\textbf{VWAP} & 5 min & USD & \begin{tabular}[c]{@{}c@{}}Average price of a security within \\ a day,  adjusted for its volume.\\  Available for an API call\\  only for traditional stocks; \\ manually calculated using \\ the formula for crypto.\end{tabular} \\ \hline
\textbf{Ticker} & - & - & \begin{tabular}[c]{@{}c@{}}Stock symbol (e.g., TSLA, AAPL \\ for traditional, BTC, ETH for crypto)\end{tabular} \\ \hline
\textbf{Interval} & 5 min & min/hr/day & Time difference between two data points \\ \hline
\end{tabular}%
}
\caption{Volume Weighted Average Price: Data \\
Data Source: Alpha Vantage API}
\label{table:vwap}
\end{table}

\subsection{Results}
\subsubsection{Moving Average Crossover}

Here, we present one typical cryptocurrency, Ether (ETH), to generate buy-and-sell signals, followed by the moving average crossover strategies. We visualize our holdings and cash flow during this testing period, and we determine and compare the performance of the moving average crossover to the simple buy-and-hold strategy.

\begin{figure}[!htpb]
\centering
\includegraphics[width=3.6in]{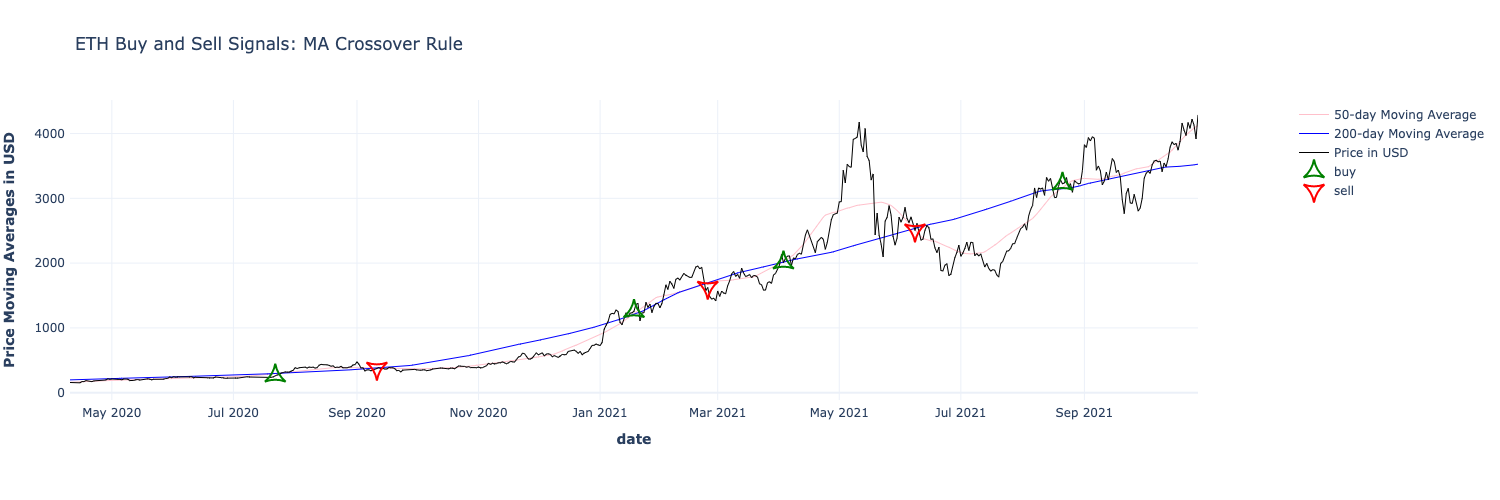}
\caption{\textbf{Buy-and-Sell Signal: ETH moving average crossover}}

\label{fig:3_1_a}
\end{figure}

Figure~\ref{fig:3_1_a} is a visualization of how the buy-and-sell strategy is defined when applying the moving average strategy to perform the backtesting on ETH. According to Figure~\ref{fig:2_1}, when the short-window moving average (SWMA) crosses above the long-window moving average (LWMA), it generates a buy signal. Conversely, it produces a sell signal. The signal is specifically drawn on the closing price line.

\begin{figure}[!htbp]
\centering
\includegraphics[width=3.6in]{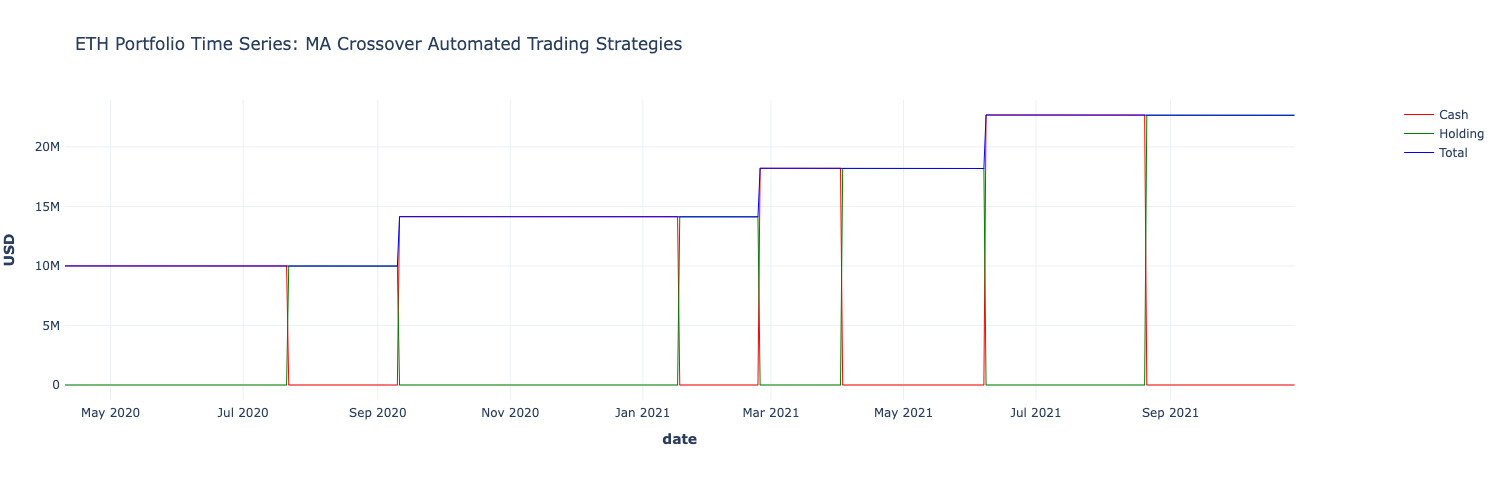}
\caption{\textbf{Portfolio time series: ETH moving average crossover}}

\label{fig:3_1_b}
\end{figure}

Figure~\ref{fig:3_1_b} shows how the portfolio would change in the time series after applying the crossover strategy to claim to buy or sell ETH during the backtesting period. Here, we assume that the transaction fee is 0.1\% per transaction. We carry out buying behavior by buying the maximum number of shares with all the cash held, and we carry out selling behavior by selling all the shares to obtain cash.

\begin{figure}[!htbp]
\centering
\includegraphics[width=3.6in]{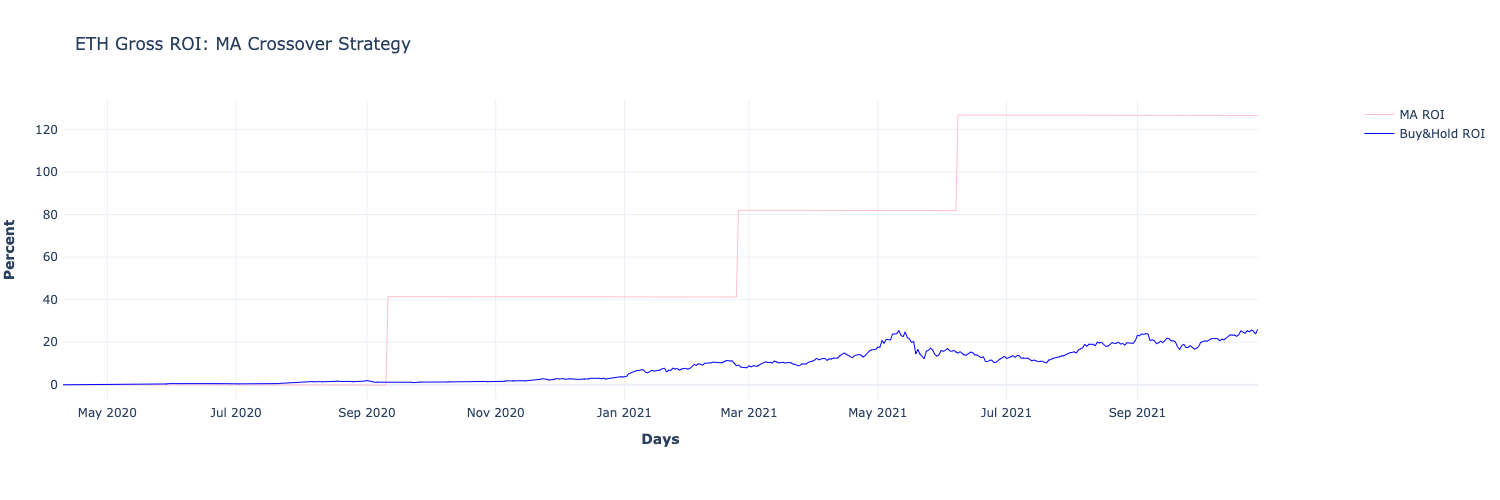}
\caption{\textbf{Gross ROI: ETH moving average crossover vs. buy-and-hold} \\
Moving average crossover strategy ROI: 849.84\%\\Buy \& hold strategy ROI: 11.97\%}

\label{fig:3_1_c}
\end{figure}

From Figure~\ref{fig:3_1_c}, it is obvious that over the two-year backtesting period, the total revenue increases.

From Figure~\ref{fig:3_1_c}, the ROI given this backtesting period is 849.8\%, performing much better than the simple buy-and-hold strategy, which produces an ROI of 12.0\%.

\begin{figure}[!htbp]
\centering
\includegraphics[width=3.6in]{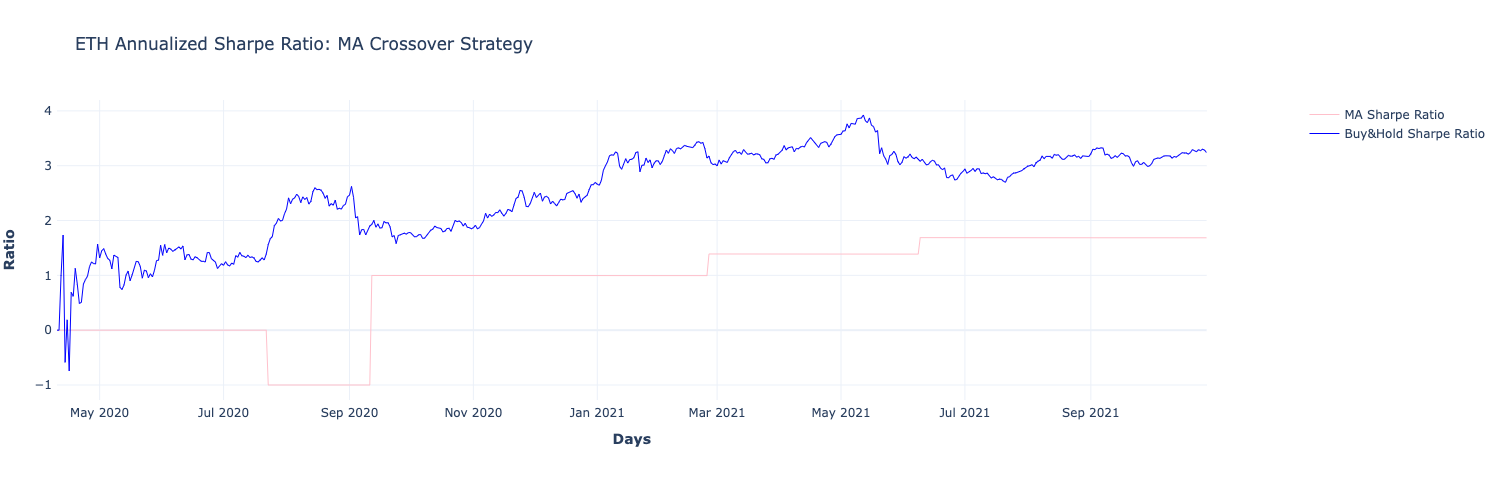}
\caption{\textbf{Sharpe Ratio: ETH moving average crossover vs. buy-and-hold} \\
Moving average crossover strategy Sharpe ratio: 0.98 \\Buy-and-hold strategy Sharpe ratio: 2.60}

\label{fig:3_1_d}
\end{figure}

From Figure~\ref{fig:3_1_d}, the Sharpe ratio given this backtesting period is $1.69$, showing that this strategy, which produces a Sharpe Ratio result of $3.24$, has a higher ability to deal with financial risks.

\subsubsection{Volume-Weighted Average Price}

\begin{figure}[!htbp]
\centering
\includegraphics[width=3.6in]{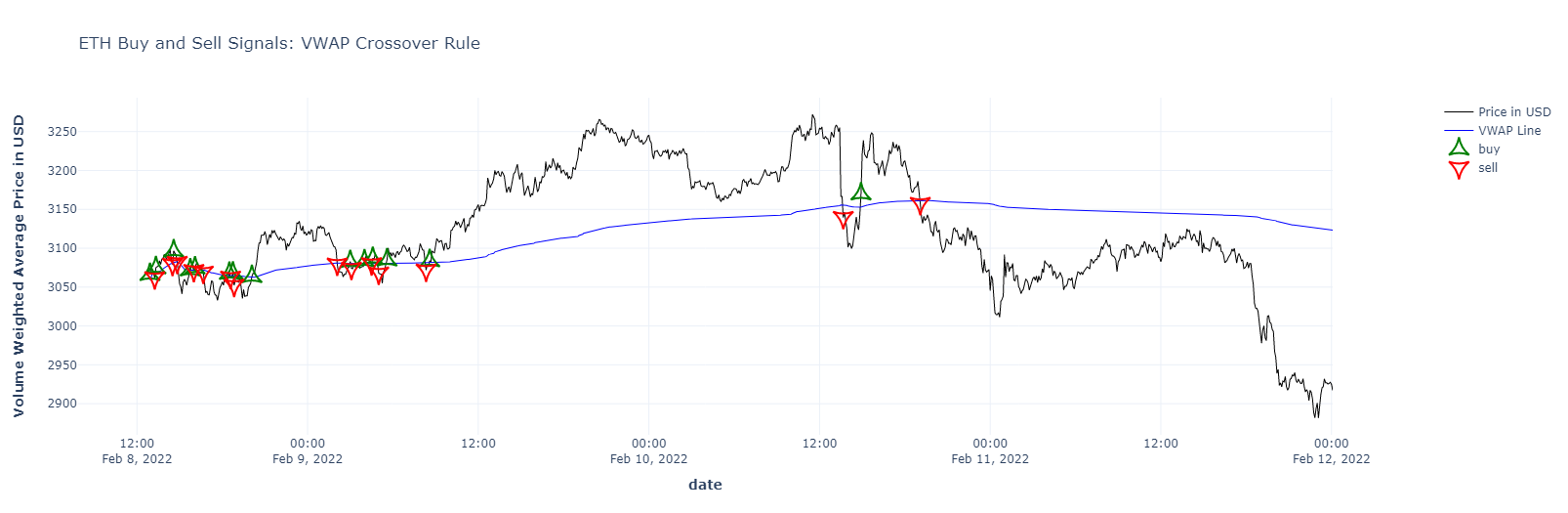}
\caption{\textbf{Buy-and-Sell Signal: Volume-Weighted Average Price}}
\label{fig:3_2_a}
\end{figure}

Figure~\ref{fig:3_2_a} shows the buy-and-sell signals generated using the VWAP strategy when trading ETH. The black line shows the evolution of the price of ETH for the period retrieved. The blue line shows the VWAP value over the same period. The value of the VWAP was calculated for each 5-min interval from the price and closing data. We see that there is an upward green arrow signifying a buy signal each time the price of ETH crosses the VWAP line from below. We have downward red arrows for sell signals, which are generated whenever the price of ETH crosses the VWAP line from above. We see that when the current price and the VWAP line intersect repeatedly in a short time period, there are multiple consecutive buy-and-sell signals.

\begin{figure}[!htbp]
\centering
\includegraphics[width=3.6in]{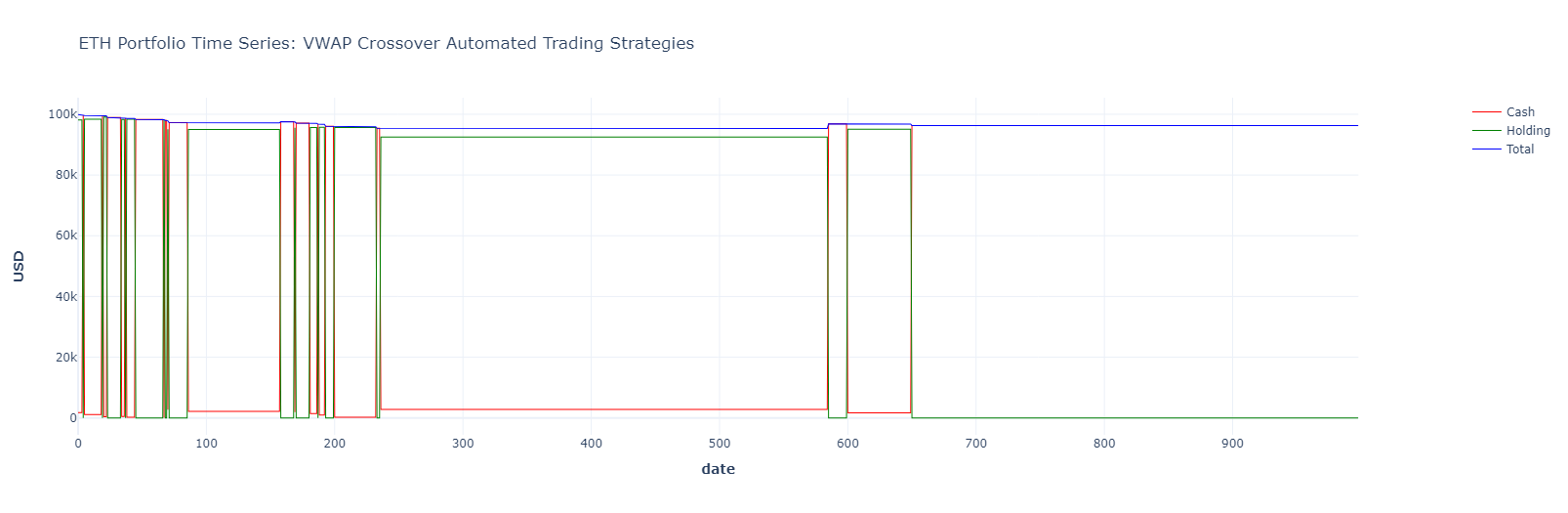}
\caption{\textbf{Portfolio time series: Volume-Weighted Average Price}}

\label{fig:3_2_b}
\end{figure}

Figure ~\ref{fig:3_2_b} displays the evolution of the portfolio when trading ETH using the VWAP strategy. We plotted the current amount of cash in red, the value of the holdings in green, and the sum of both cash and holdings in red (total). As expected from the multiple buy-and-sell signals in Figure \ref{fig:3_2_a} early on, there are many fluctuations in the amount of cash and holdings on the left side of the graph. There are far fewer as we move to the right (as there are fewer signals produced there). The overall total seems to have decreased from its initial value of 100,000.

\begin{figure}[!htbp]
\centering
\includegraphics[width=3.6in]{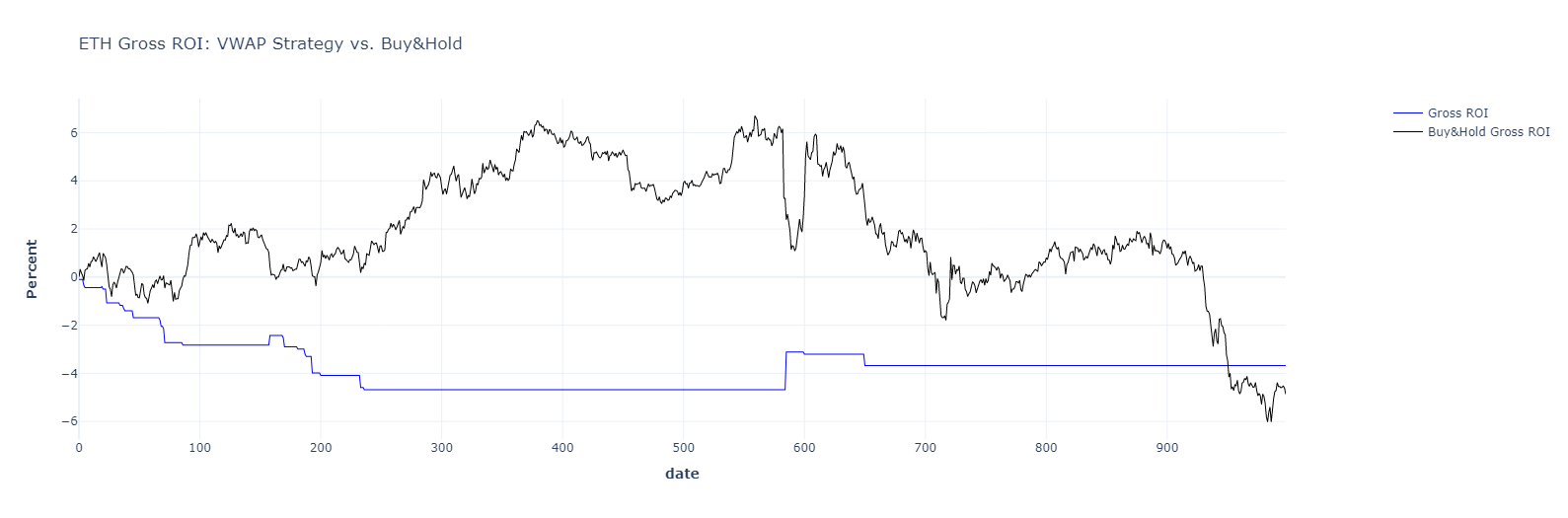}
\caption{\textbf{Gross ROI: Volume-Weighted Average Price vs. Buy-and-Hold} \\
Volume-weighted average price strategy ROI: -3.93\%\\Buy-and-hold strategy ROI: -4.26\%}

\label{fig:3_2_c}
\end{figure}

Figure ~\ref{fig:3_2_c} displays a comparison of the ROI when trading ETH using the buy-and-hold strategy (in black) and the VWAP strategy (in blue). The buy-and-hold strategy looks the same as the price evolution graph in Figure \ref{fig:3_2_a}. For the studied time period, the trader would have an ROI of approximately -4\%. Using the VWAP also results in a negative ROI, very close to the buy-and-sell ROI. For this time period, using the buy-and-sell strategy or the VWAP strategy would have resulted in a loss.

\begin{figure}[!ht]
\centering
\includegraphics[width=3.6in]{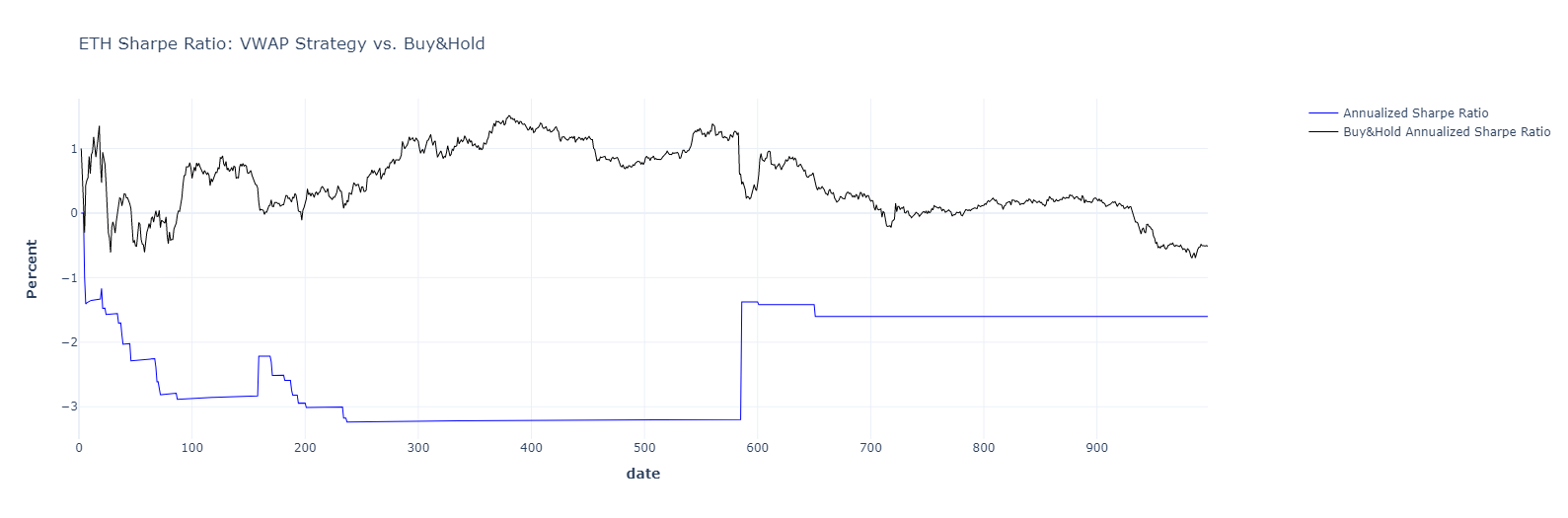}
\caption{\textbf{Sharpe Ratio: Volume-Weighted Average Price vs Buy-and-Hold} \\
Volume-Weighted Average strategy Sharpe ratio: -1.62 \\Buy-and-hold strategy Sharpe ratio: -0.52}

\label{fig:3_2_d}
\end{figure}

Figure ~\ref{fig:3_2_d} displays a comparison of the Sharpe ratio when trading ETH using the buy-and-hold strategy (in black) and the VWAP strategy (in blue). We see that both Sharpe ratios end up being negative, with the buy-and-hold ratio being slightly below zero, and the VWAP ratio being close to -2\%. Both the buy-and-hold and VWAP strategies return negative Sharpe ratios in this case. We therefore cannot say that using the VWAP improves our returns, especially in the very short term.

\section{Discussion}
In conclusion, we propose a data science pipeline for algorithmic trading and verify that the pipeline is generally applicable for designing, programming, and evaluating the algorithmic trading of stock and crypto assets. Furthermore, we implement our algorithms through object-oriented programming in Python3, which serves as open-source software for future academic research and applications. The data, code, and supplementary results can be found on the Github repository: \url{https://github.com/SciEcon/SRS2021}.

We evaluated our data science pipeline based on two conventional algorithmic trading strategies. 
\begin{itemize}
    \item \textbf{The moving average crossover} is among the most prominent investment strategies in finance that build mathematical models to further forecast future market movements through historical market records~\cite{brown_2019_moving}. Currently, researchers and industry practitioners are keenly interested in analyzing the moving average crossover strategy in crypto trading. \citet{fang_2022_cryptocurrency} applied the simple moving average trading strategy based on daily closing price data for the 11 most traded cryptocurrencies over the period 2016-2018. Their results showed that technical trading rules, excluding Bitcoin, produced a desirable annualized excess return of 8.76\%. Moreover, due to the high volatility of crypto prices, a revised approach to the moving average has started to be introduced by considering the tolerance of time ranges for the intersection of transaction signals in real industry practices~\cite{interdax_2019_research}.
    \item \textbf{The volume-weighted average price} can be viewed as an advanced version of the \textit{moving average crossover strategy} that averages by weights of transaction volumes. Although the prior literature has mainly focused on using the VWAP as one of the technical indicators used to trade assets, few efforts have been made to use the index for cryptocurrencies. One emerging trend, however, is the adoption of the VWAP in the industry practices of DeFi applications. A prominent example is the Ampleforth DeFi protocol~\cite{kuo2019ampleforth}, which issues stable coin AMPL. Ampleforth uses VWAP data from decentralized Chainlink oracles~\cite{kuo_2020_the} to expand or contract the supply of AMPL accordingly, every 24 hours, adjusting the price of AMPL to be approximately one US dollar regardless of the volatility in the market demand for AMPL~\cite{zhang2021optimal}.  
\end{itemize}
\par
We also provide an online appendix in the Github repository:\url{https://github.com/SciEcon/SRS2021}. 
In the \href{https://github.com/SciEcon/SRS2021/tree/main/More\%20about\%20the\%20paper}{online appendix}, we present additional results of the aforementioned two algorithms, the moving average crossover and volume-weighted average price, and we provide an additional evaluation of our data science pipeline based on two additional algorithms: \textbf{sentiment analysis} and \textbf{pairs trading}.

Our research is seminal in inspiring future innovations in two ways.
\begin{itemize}
    \item First, the current research shows how our data science pipeline is generally applicable for implementing and evaluating existing trading algorithms. Furthermore, our approach can be applied to design, implement, and evaluate new trading algorithms, as in~\citet{liu2022cryptocurrency}. Since there currently does not exist a consensus on the valuation of crypto assets, effective algorithmic trading strategies for various crypto tokens have yet to be found.
    \item Second, our data science pipeline is general enough for researchers to customize various settings, such as data sources, input variables, assets, trading conditions, indices, buy-and-sell signals, and evaluation indicators. However, the pipeline might better serve more advanced reinforcement learning algorithms and sentiment analysis by incorporating existing general-purpose or domain-specific data science frameworks, as in~\cite{cryptoEnv2022}.
\end{itemize}

\section*{Acknowledgments}

We thank the Alpha Vantage API for providing academic accounts for the data querying in our research. We thank Zesen Zhuang for his assistance in hosting tutorial sessions for Tianyu Wu, Saad Lahrichi, Carlos-Gustavo Salas-Flores, and Jiayi Li about object-oriented programming in Python3. 

\IEEEtriggercmd{\enlargethispage{-5in}}
\bibliographystyle{IEEEtranN}
\bibliography{SRS_paper}

\end{document}